\documentclass[%
 reprint,
 superscriptaddress,
 amsmath,amssymb,
 aps,
prapplied,
floatfix,
]{revtex4-2}
\usepackage{graphicx}
\usepackage{dcolumn}
\usepackage{bm}
\usepackage{siunitx}
\usepackage{sidecap}
\usepackage[columnwise]{lineno}

\usepackage{hyperref}

\begin{document}
\preprint{}
\title{Automated extraction of capacitive coupling for quantum dot systems}

\author{Joshua Ziegler}
\altaffiliation{Current address: Intel Components Research, Intel Corporation, 2501 NW 229th Avenue, Hillsboro, Oregon 97124, USA}
\affiliation{National Institute of Standards and Technology, Gaithersburg, Maryland 20899, USA}

\author{Florian Luthi}
\affiliation{Intel Components Research, Intel Corporation, 2501 NW 229th Avenue, Hillsboro, Oregon 97124, USA}

\author{Mick Ramsey}
\affiliation{Intel Components Research, Intel Corporation, 2501 NW 229th Avenue, Hillsboro, Oregon 97124, USA}

\author{Felix Borjans}
\affiliation{Intel Components Research, Intel Corporation, 2501 NW 229th Avenue, Hillsboro, Oregon 97124, USA}

\author{Guoji Zheng}
\affiliation{Intel Components Research, Intel Corporation, 2501 NW 229th Avenue, Hillsboro, Oregon 97124, USA}

\author{Justyna P. Zwolak}
\email{jpzwolak@nist.gov}
\affiliation{National Institute of Standards and Technology, Gaithersburg, Maryland 20899, USA}
\affiliation{Joint Center for Quantum Information and Computer Science, University of Maryland, College Park, Maryland 20742, USA}

\date{\today}
\begin{abstract}
Gate-defined quantum dots (QDs) have appealing attributes as a quantum computing platform. 
However, near-term devices possess a range of possible imperfections that need to be accounted for during the tuning and operation of QD devices. 
One such problem is the capacitive cross-talk between the metallic gates that define and control QD qubits.
A way to compensate for the capacitive cross-talk and enable targeted control of specific QDs independent of coupling is by the use of virtual gates.
Here, we demonstrate a reliable automated capacitive coupling identification method that combines machine learning with traditional fitting to take advantage of the desirable properties of each.
We also show how the cross-capacitance measurement may be used for the identification of spurious QDs sometimes formed during tuning experimental devices. 
Our systems can autonomously flag devices with spurious dots near the operating regime, which is crucial information for reliable tuning to a regime suitable for qubit operations.
\end{abstract}

\maketitle
\section{Introduction}
Quantum dot (QD) arrays, in which charge carriers are trapped in localized potential wells and qubits can be made by use of the spin and permutation symmetries of the carriers, are a promising quantum computing platform~\cite{Pillarisetty21-SQC, Bavdaz22-QDC, Boter22-SWA}.
In fact, the first demonstrations of QD two-qubit gates with fidelities exceeding the thresholds for fault-tolerant computing were developed in 2022~\cite{Noiri22-STL, Xue22-QLS, Mills22-TSP}.
However, because the individual charge carriers that make up qubits have electrochemical sensitivity to minor impurities and imperfections, calibration and tuning of QD devices is a nontrivial and time-consuming process, with each QD requiring careful adjustment of a gate voltage to define charge number, and multiple gate voltages to specify tunnel coupling between QDs for two-qubit gates or to reservoirs for reset and measurement.
While manual calibration is achievable for small, few-QD devices, with increasing size and complexity of QD arrays the relevant control parameter space grows quickly, necessitating the development of autonomous tuning methods. 

There have been numerous demonstrations of automation of the various phases of the tuning process for single- and double-QD devices~\cite{Zwolak21-AAQ}. 
Some approaches seek to tackle tuning starting from device turn-on to coarse tuning~\cite{Baart16-CAT, Darulova19-ATQ, Moon20-ATQ, Czischek21-MNA}, while others assume that bootstrapping (calibration of measurement devices and identification of a nominal regime for further investigation) and basic tuning (confirmation of controllability and device characteristics) have been completed and focus on a more targeted automation of the coarse and charge tuning~\cite{Zwolak18-QLD, Durrer19-ATQ, Zwolak20-AQD, Zwolak21-RBI, Lapointe-Major19-ATQ}. 
While the initial autotuning approaches relied mainly on the appealingly intuitive and relatively easy-to-implement conventional algorithms that typically involved a combination of techniques from regression analysis, pattern matching, and quantum control theory, the more recent algorithms take advantage of the modern computer vision and machine learning~\cite{Zwolak21-AAQ}. 

A typical accumulation-mode QD device consists of two sets of gates---plungers and barriers---that collectively control the overall potential profile, the QD-specific single-particle energy detuning of individual QDs, the tunnel couplings between QDs, and tunnel rates between the outermost QDs and reservoirs.
Ideally, each plunger gate would affect only the electrochemical potential of a single targeted QD and each barrier gate only one intended tunnel barrier.
Due to the tight proximity, however, each gate capacitively couples to nearby potential and tunnel barriers.
This makes careful control of these key parameters challenging.

One way to compensate for the capacitive crosstalk between gates is to enable orthogonal control of the QDs potential by implementing so-called \emph{virtual gates}~\cite{Hensgens17-FHQ}.
Specifically, linear combinations of gate voltage changes can be mapped onto onsite energy differences~\cite{Oosterkamp98-MSQ, Hensgens17-FHQ, Hensgens18-PhD, Perron15-QSB}.
These approaches have been key for the initialization and control of larger QD arrays~\cite{Volk19-LQR, Mills19-SSC}. 

To autonomously identify capacitive couplings in a device, various approaches have been demonstrated using both conventional fitting and machine learning (ML) techniques~\cite{vanDiepen18-ATC, Mills19-CAT, Oakes20-AVV, Liu22-ACT}.
However, these approaches, typically relying on the Hough transform or conventional least-squares fitting procedures, may be unreliable in the presence of data imperfections.
Hough transforms can extract slopes directly but may be sensitive to noise or be excessively complex to analyze. 
Conventional fitting can be more flexible but is susceptible to local minima and can be time-consuming at inference time.

Convolutional neural networks (CNNs) are well suited for extracting high-level features from images and can remain effective in the presence of noise or other imperfections~\cite{Ziegler22-TRA}.
However, ML methods can have difficulties identifying data outside of the training distribution even if they contain similar features~\cite{Darulova20-EDM}.
Fortunately, given a simplified, high-level representation of the data, conventional fitting approaches can be more targeted to extract key information more effectively and quickly.

Here we develop a reliable automated capacitive coupling identification method that combines ML with traditional fitting to take advantage of the desirable properties of each.
We use an ML module for pixel classification followed by linear regression for extracting targeted information and demonstrate effective performance across varying noise levels and data variations.
Testing each of these methods on a set of eight simulated QD devices with large variability and realistic noise variation mimicking experimental conditions shows that the approach combining ML and traditional fitting works well, with a root-mean-square error (RMSE) of $0.034(14)$, corresponding to a roughly $8~\%$ error, for predicting virtual gate matrix off-diagonal values (normalizing such that diagonal values are one)~\footnote{We use a value(uncertainty) notation to express uncertainties, for example, $1.5(6)~\si{\centi\meter}$ would be interpreted as $(1.5\,\pm\,0.6)~\si{\centi\meter}$. 
All uncertainties herein reflect the uncorrelated combination of single-standard deviation statistical and systematic uncertainties.}.
This RMSE roughly corresponds to the error rate expected for previous cross-capacitance extraction methods~\cite{Mills19-CAT} that required multiple iterations and higher data quality.

We also demonstrate how the cross-capacitance measurement may be used for the identification of spurious QDs formed during tuning experimental devices.
Many of the autotuning approaches proposed to date rely on a series of small two-dimensional (2D) scans capturing a relatively narrow range of the voltage space~\cite{Zwolak20-AQD, Durrer19-ATQ, Ziegler22-TRA, Ziegler22-TAR}.
While such approaches improve the efficiency of tuning, they may result in unexpected and difficult-to-assess failure modes when the tuning algorithm terminates at an anticrossing with a spurious QD that may form in small potential wells due to interface defects, surface roughness, or strain within the device~\cite{Thorbeck12-UQD}.
They are highly undesirable since they may interfere with the QDs intended for use as qubits and cannot themselves be used as qubits.
To avoid device tuning failure, spurious QDs must be identified when present and avoided.
We test the utility of our approach for capacitive coupling estimation by identifying spurious QDs in experimental measurements of QD devices~\cite{Pillarisetty21-SQC}.

This paper is organized as follows: In Sec.~\ref{sec:methods} we introduce the framework of combining traditional fitting techniques with a pixel classifier to process the high-level information extracted from experimental data.
In Sec.~\ref{sec:results} we show the utility of the proposed framework to automatically extract virtual gates as well as identify charge transitions resulting from a formation of spurious QDs. 
Finally, in Sec.~\ref{sec:conclusion} we summarize the results and discuss the outlook.

\section{Methods: Machine learning and fit}
\label{sec:methods}
Capacitive couplings in a QD device can be measured and, in a constant capacitance approximation, described by a matrix that maps the physical gate voltages onto the effect they each have on the QD's chemical potentials or barriers~\cite{Keller96-AEC, Hensgens17-FHQ, vanDiepen18-ATC, Mills19-CAT, Hsiao20-EOT, Qiao20-CME}.
Measurement of the elements of this matrix must be performed distinctly for electrochemical potentials and tunnel barriers.
Couplings of the chemical potentials to each QD---the focus of this work---can be extracted from shifts in charge transition lines when each voltage is varied~\cite{Hensgens17-FHQ}, while the effect of each gate on tunnel barriers can be assessed by measuring changes in the width of interdot transitions, assuming the electron temperature is sufficiently low~\cite{Hsiao20-EOT}.
Measured this way, the couplings are relative, usually scaled with respect to the coupling of the QD to the nearest gate.
An absolute energy scale can be obtained by measuring the gate lever arms with photon-assisted tunneling, Coulomb diamonds, or bias triangles~\cite{Hanson07-SQD}.
However, for establishing the orthogonal control the relative scale is sufficient~\cite{Volk19-LQR}.

For a double QD, the virtualization matrix relating the physical plunger gates to virtual gates can be represented by the equation
\begin{equation}
\label{eq:vg_dd_submatrix}
    \begin{pmatrix}
        V_{P'_1} \\
        V_{P'_2} \\
    \end{pmatrix}
    = 
    \begin{pmatrix}
        1           & \alpha_{12} \\
        \alpha_{21} & 1 \\
    \end{pmatrix}
    \begin{pmatrix}
        V_{P_1} \\
        V_{P_2} \\
    \end{pmatrix}
\end{equation}
Each row is normalized such that the diagonal entries are $1$ to reflect the relative nature of our virtual gates.

\begin{figure}[t]
    \centering
    \includegraphics[width=\linewidth]{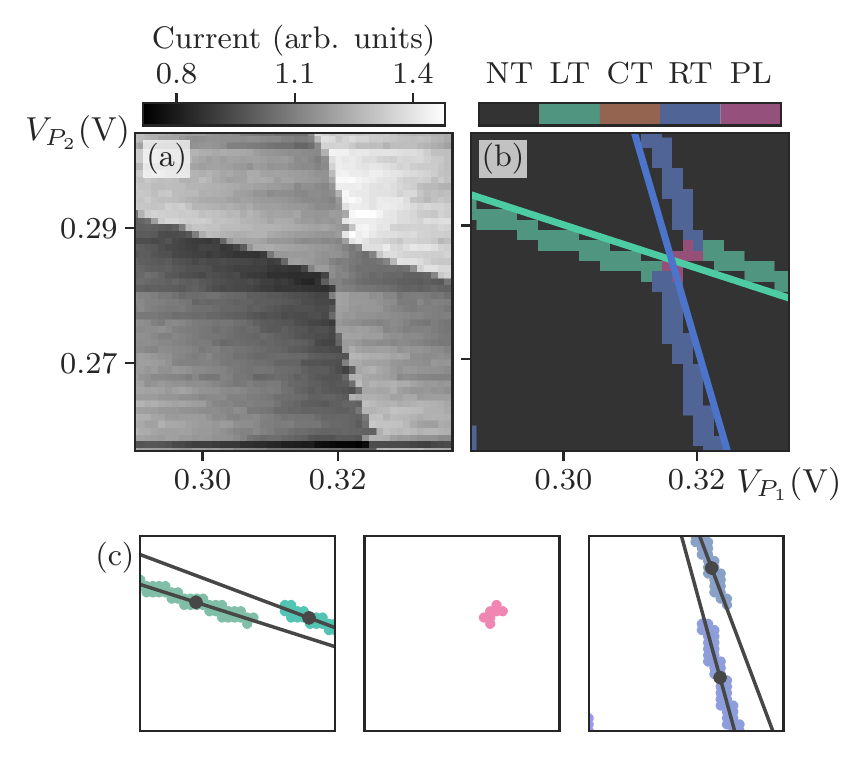}
    \caption{An example 2D scan and corresponding pixel classification, class clusters, and linear fits. 
    (a) A simulated voltage scan showing left and right transitions as well as a polarization line. 
    (b) Pixel classification for the scan shown in (a). 
    (c) Regions of pixels and linear fits from the pixel classification. 
    The large dark points indicate the centers of pixel regions.}
    \label{fig:sim_sub_example}
\end{figure}

The relative cross-capacitances for chemical potentials manifest themselves via the slopes of charge transition lines, with the dominant terms of the cross-capacitance matrix determined from a measurement in the space of neighboring pairs of gates~\cite{Volk19-LQR}.
We address the identification of the cross-capacitances as captured in 2D plunger-plunger gate scans, as shown in Fig.~\ref{fig:sim_sub_example}(a).
To translate the low-level QD data into high-level information useful for automation we use a {\it pixel classifier}, that is, a CNN model with a structure similar to a feature pyramid network~\cite{Lin17-FPN}.
Additional details about the CNN design can be found in Appendix~\ref{app:cnn}.
The pixel classifier takes as an input a small 2D plunger voltage scan obtained using a charge sensor, as shown in Fig.~\ref{fig:sim_sub_example}(a).
It then identifies each pixel within the scan as belonging to one of the charge transition classes---that of left QD, right QD, central QD, or interdot (polarization line) transition, denoted by LT, RT, CT, or PL, respectively---or to the no transition (NT) class.
In other words, the CNN provides a high-level classification of the raw experimental data while keeping spatial information about the relative location and orientation of the detected features, which is useful for algorithmic processing.
Figure~\ref{fig:sim_sub_example}(b) shows the pixel classification of a scan from Fig.~\ref{fig:sim_sub_example}(a).

To translate pixel classifications into capacitive couplings, we identify contiguous regions within each class of pixels in an image and then independently fit them using a linear regression model.
A labeling algorithm from the multidimensional image processing package in SciPy is then used to determine the relevant clusters of connected pixels for each class~\cite{SciPy}.
This separates charge transitions into distinct lines identified by an index and an assigned class so that each can be processed individually.
The $x$ and $y$ pixel indices of each region of pixels classified as LT, CT, or RT are independently fitted using linear regression, as shown in Fig.~\ref{fig:sim_sub_example}(c).
When multiple segments for a given class are present in an image, the capacitive coupling returned is the average for all fitted lines weighted by the number of pixels in each cluster and the standard deviations of the respective fits, yielding the solid lines in Fig.~\ref{fig:sim_sub_example}(b) (offset arbitrarily for comparison with the pixel regions).
To facilitate a more direct calculation of the fit error for the RT capacitive coupling, the $x$ and $y$ indices are inverted before linear regression.
Standard deviations $\sigma$ are computed from the standard error of the fit, $S$, by $\sigma = S/\sqrt{n}$, where $n$ is the number of pixels in the pixel region, as in Student's $t$-distribution~\cite{Student08-PEM}.
In addition, each region is tagged with its center in voltage space, shown by the large black points in Fig.~\ref{fig:sim_sub_example}(c), which allows tracking of the changes in charge transitions and their slopes within the larger space.

\subsection{Data}
\label{ssec:data}
The data used for training the ML tools and testing the methods were generated using a simulation of QD devices~\cite{Zwolak18-QLD}.
The simulation is composed of a calculation of the electron density in the Thomas-Fermi approximation and a capacitance matrix to determine the stable electron configuration.
To improve the robustness of the models, the data are augmented with synthetic white, pink ($1/f$), and telegraph noise~\cite{Ziegler22-TRA}.
The effect of a QD charge sensor strongly coupled to the plunger gates is varied during the scan to improve performance on this type of experimental data.

\begin{figure}[t]
    \centering
    \includegraphics[width=\linewidth]{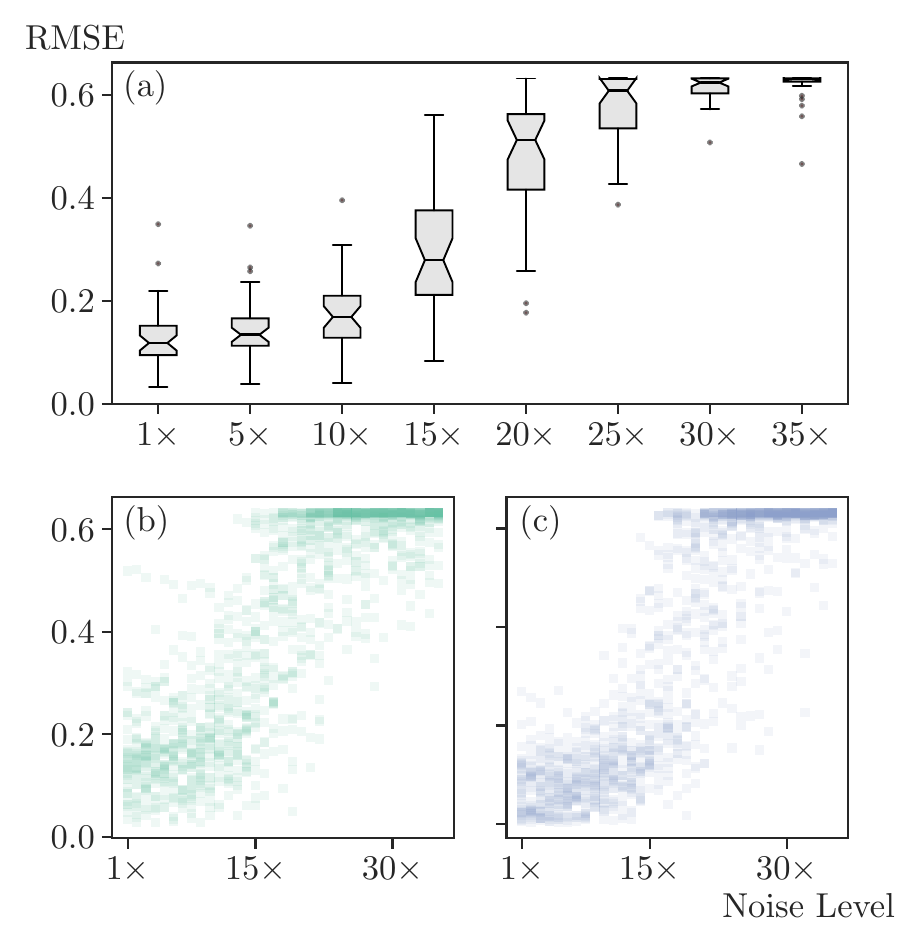}
    \caption{(a) Box plot of root-mean-square error (RMSE) for all transition classes (left, central, and right [LT, CT, RT]) as a function of the synthetic noise level.
    The notch indicates the $95~\%$ confidence level.
    (b) RMSE as a function of noise level for the LT class. 
    (c) RMSE as a function of noise level for the RT class.}
    \label{fig:rmse_pix_class}
\end{figure}

The training dataset consists of $1.6\times10^5$ devices with parameters varied over a uniform distribution with a standard deviation equal to $1~\%$ of each parameter's value.
To train the ML models we randomly sample 10 small scans per device and use charge-state ground truth to label each scan at a pixel level with the presence and type of transition, yielding NT, LT, CT, RT, and PL labels.
Additionally, we extract the slopes of the transition lines directly using the gradients of the device charge.

\begin{figure*}
    \centering
    \includegraphics[width=\linewidth]{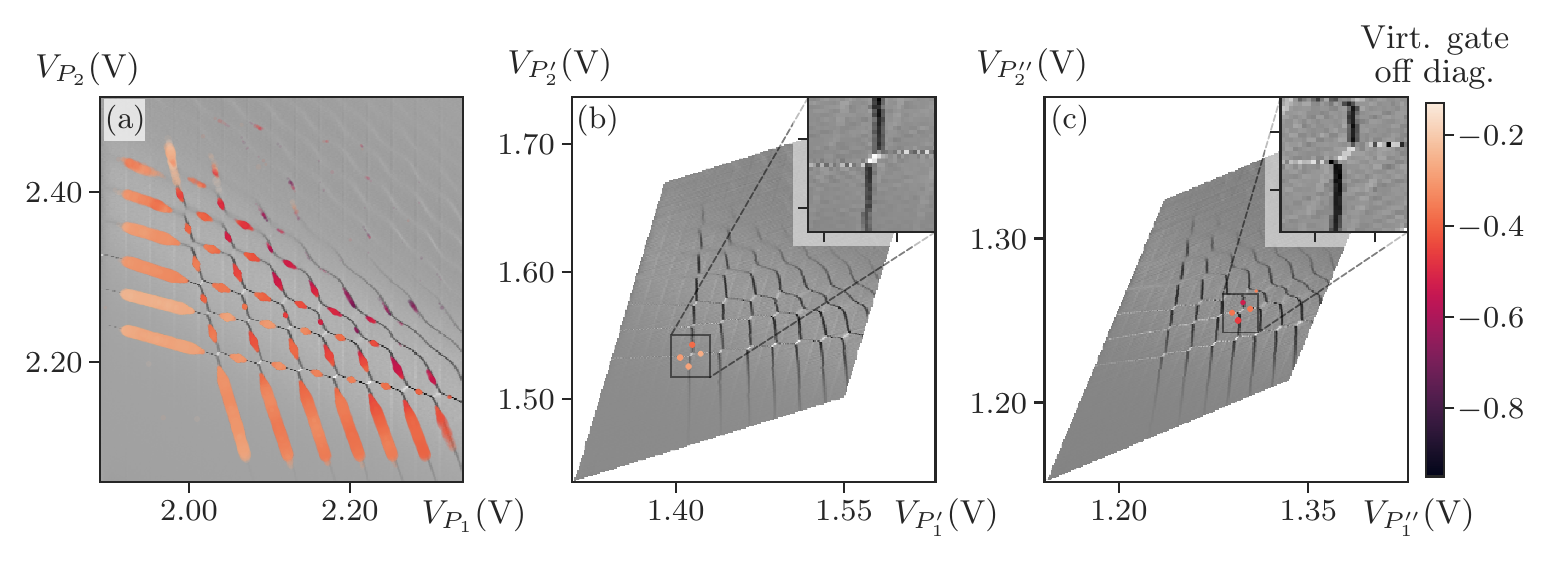}
    \caption{(a) Large experimentally measured charge stability diagram with a scatter plot of centers of pixel class regions overlaid. 
    The colors of the points indicate the virtual gate off-diagonal values identified by fits to the region. 
    The sizes of the points indicate the weights used when averaging. Only points with relative error less than $20~\%$ are plotted. 
    (b),(c) Charge stability diagram after applying virtual gates acquired near the $(0,0)$--$(1,1)$ charge transition in (b) and near the $(1,3)$--$(2,4)$ charge transition in (c).
    In both (b) and (c) the virtualization is performed off-line, via an affine transform to the original scan shown in (a), and the points are plotted using the same parameters as in (a).
    }
    \label{fig:slope_affine_transformed_maps}
\end{figure*}

The test data consist of eight simulated devices with large variations in screening length and device pitch and with large shifts in the position of one of the plunger gates.
These changes lead to large variations in the slopes of and spacing between the charge transition lines, the capacitive coupling between QDs, and the relative sizes of left and right QD regions, making them largely distinct from the training data.
To facilitate a controlled study and track the performance of the pixel classifier as data quality degrades, each large scan is randomly sampled $50$ times and the resulting small scans are augmented with increasing levels of synthetic noise. 
This results in a set of 400 simulated test scans. 
Finally, several large experimental measurements acquired using a double-QD configuration on a three-QD Si$_x$/SiGe$_{1-x}$ device, fabricated on an industrial 300-\si{\milli\meter} process line~\cite{Pillarisetty21-SQC}, are used to test the performance of the virtualization algorithm.
Experimental scans capturing spurious QDs are used to demonstrate the algorithm for spurious QD detection.

\section{Results}\label{sec:results}
We test the effectiveness of our automated approach to extracting the cross-capacitance by first evaluating the performance of each component (i.e., the pixel classifier and the slope extractions) on each scan in the simulated test set.
The error of the pixel classifier in our framework is defined as a fraction of pixels belonging to true transitions that are not contained in line segments in the CNN output.
This captures type II errors (false negatives) without the effect of false type I errors (false positives) due to imperfect labels~\footnote{Calculating the error of the pixel classifier directly is made difficult by necessary but imperfect resizing of labels on evaluation.}.
Figure~\ref{fig:rmse_pix_class}(a) shows the change in RMSE of the extracted off-diagonal virtual gate values as a function of the noise level in the simulated data. 
At the noise level of $1.0$ (i.e., the noise level estimated from experimental data in Ref.~\cite{Ziegler22-TAR}) we observe an RMSE of $0.17(5)$.
The RMSE increases significantly to $0.50(11)$ at the noise level of $20$.
For reference, a pixel classifier that always predicts the NT class would have an RMSE of $0.62$ ($\sqrt{0.4}$).
For the LT and RT classes relevant to cross-capacitances, shown in Figs.~\ref{fig:rmse_pix_class}(b) and \ref{fig:rmse_pix_class}(c), the pixel classifier for noise level $1.0$ has an RMSE of $0.20(8)$ and $0.11(8)$, respectively.

To verify that the slope extraction tool works as intended, we test it across the eight large simulated test devices.
For these tests, we evaluate the pixel classifier in windows of size roughly $1.5$ times the charging energy, as estimated by the spacing of the first two charge transitions. 
Outputs from the pixel classifier are cropped by one pixel from the edge of the image before processing because the zero padding at each layer causes reduced performance at image edges~\cite{Ronneberger15-UNET}.
The resulting classes of pixels are then grouped into distinct clusters. 
For each cluster consisting of more than five pixels an independent linear fit is performed, returning both the slope and the standard error of the fitted line.
The RMSE for the linear fits compared to the cluster from pixel classifications is $0.6$ pixels, which is near the minimum relative to the two-pixel width of lines in pixel classifications.
This information can be used to find the orthogonal ``virtual'' control space or to flag transitions that potentially belong to spurious dots, as described in the following subsections.

\subsection{Deriving virtual gates}\label{ssec:vg}
As stated in Sec.~\ref{sec:methods}, we derive the off-diagonal elements of the capacitance matrix (defining the virtual gates transformation) based on the slopes of the LT and RT captured in a given image, while the diagonal elements are set $1.0$.
When multiple lines belonging to the same class are detected, as in Fig.~\ref{fig:sim_sub_example}(a), the capacitive coupling is calculated through a weighted average~\cite{Student08-PEM}.

The off-diagonal elements of the virtualization matrix computed this way have an RMSE of $0.034(14)$ at the noise level of $1.0$ defined in Ref.~\cite{Ziegler22-TAR}, corresponding to a roughly $8~\%$ error compared to the ground-truth values derived from simulated data.
We further test them on a range of levels of synthetic noise and find the RMSE rises by a factor of two at a level of noise roughly 
$15$ times the level of noise defined in Ref.~\cite{Ziegler22-TAR}, consistent with the pixel classifier error.

To better understand the trends of the virtualization matrix in the plunger-plunger space, we carry out a performance analysis using the simulated test set and several experimentally measured scans.
For each scan, we calculate the fits to the pixel classification clusters based on a series of small scans sampled at each point within the large scan with the exclusion of a margin implemented to ensure that all sampled scans fall within the full scan boundaries.
The small scans and the margins are set to have a size $1.5$ times the charging energy of a given simulated device.
Figure~\ref{fig:slope_affine_transformed_maps}(a) shows the centers of the pixel region identified in each small scan [as in Fig.~\ref{fig:sim_sub_example}(c)] as the sampling window is swept across a large experimentally measured charge stability diagram.
The regions identified by the pixel classification are consistently placed correctly on the charge transition lines regardless of the position of the line within a small scan.
Region centers shift along the charge transition lines as different portions of the line are captured within the small scan and remain fixed whenever the same fragment of the charge transition is captured.
The color of the points indicates the off-diagonal values of the virtual gate matrix, $\alpha_{12}$ and $\alpha_{21}$.
As expected, these coupling constants grow larger in magnitude as charges are added to each QD.
Finally, the size of the points in Fig.~\ref{fig:slope_affine_transformed_maps}(a) indicates the $1/\sigma^2$ weight of the slopes used when averaging multiple slopes from the same type of transition within a small scan.
As desired, the positions of the points with smaller sizes indicate that lines that are smaller or less captured within a small scan have fits with larger errors.
Overall, this plot confirms that the combination of pixel classification with the fits is working as intended at capturing charge transition lines and their slopes.

\begin{figure}
    \centering
    \includegraphics[width=\linewidth]{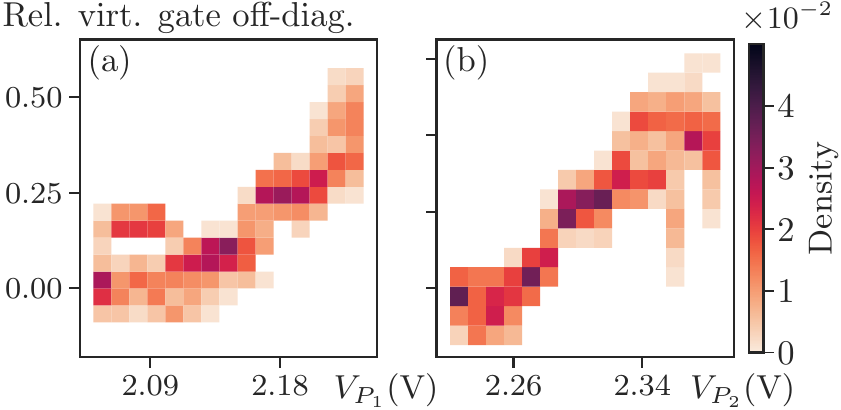}
    \caption{Histograms of the off-diagonal elements of the virtualization matrix for an experimentally measured scan shown in Fig.~\ref{fig:slope_affine_transformed_maps}(a) as a function of plunger gates (a) $V_{P_1}$ and (b) $V_{P_2}$. 
    Off-diagonal values are shifted and scaled by the mean of the off-diagonal elements from virtual gates in the $(1,1)$ charge state for ease of comparison.
    Virtual gate values are extracted from a strip of small scans shifted by 3~\si{\milli\volt} (two pixels) in each opposing direction to better visualize variation at each plunger gate value.}
    \label{fig:vg_1d_variation}
\end{figure}

To demonstrate the spatial relevance of the virtual gates derived from a set of fits across a device's charge landscape, in Figs.~\ref{fig:slope_affine_transformed_maps}(b) and \ref{fig:slope_affine_transformed_maps}(c) we plot affine-transformed charge stability diagrams, with points indicating fits overlaid.
The points plotted are the centers of pixel regions with the same color and size encoding as in (a).
The affine transformation applied in Fig.~\ref{fig:slope_affine_transformed_maps}(b) corresponds to virtual gates derived from an image near the $(0,0)$--$(1,1)$ charge transition with off-diagonal values $\alpha_{12} = -0.282(4)$ and $\alpha_{21} = -0.331(4)$.
For Fig.~\ref{fig:slope_affine_transformed_maps}(c), the affine transformation applied has virtual gates from the $(1,3)$--$(2,4)$ charge transition, with off-diagonal values $\alpha_{12} = -0.363(4)$ and $\alpha_{21} = -0.480(4)$.
As can be seen in the insets in Figs.~\ref{fig:slope_affine_transformed_maps}(b) and \ref{fig:slope_affine_transformed_maps}(c), these virtual gates are very effective at transforming the target region into an orthogonal space, but the difference between the extracted virtual gate off-diagonal values is about $50~\%$ higher for the latter case.
This highlights the importance of an efficient local method for determining virtual gates.

To further understand how capacitive coupling varies across a charge stability diagram, we can calculate variation as each plunger gate is adjusted.
Figures~\ref{fig:vg_1d_variation}(a) and \ref{fig:vg_1d_variation}(b) show how cross-capacitances extracted from small scans change as $V_{P_1}$ and $V_{P_2}$ are varied.
To better show the trend of expected values, cross-capacitances from small scans shifted by 3~\si{\milli\volt} (two pixels) in each opposing direction are included.
This shows that the cross-capacitances extracted from small scans effectively capture variation across charge stability diagrams.

\subsection{Detection of spurious dots}
\label{ssec:spur-dots}
Visually, spurious QDs are recognized in large 2D scans as charge transitions with slopes diverging from a monotonic trend.
In this framework, they may be identified as transition lines with anomalous capacitive couplings relative to the transitions around them.

\begin{figure*}
    \centering
    \includegraphics[width=\linewidth]{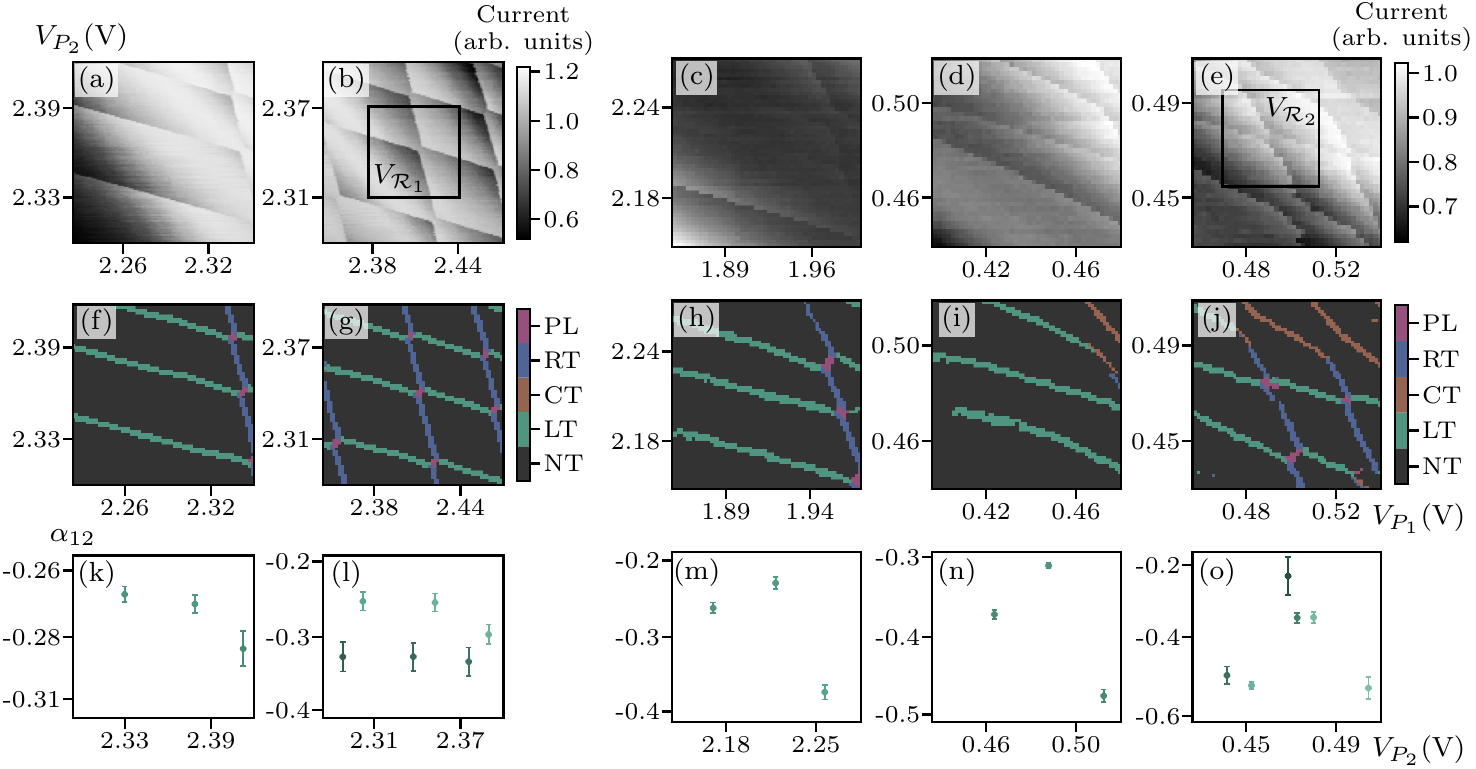}
    \caption{Spurious dot detection.
    The top row shows two charge stability diagrams capturing properly formed QD [panels (a),(b)] and three charge stability capturing spurious QD [panels (c)--(e)].
    The black boxes in (b) and (e) highlight small 2D scans, denoted by $V_{\mathcal{R}_1}$ and $V_{\mathcal{R}_2}$, typical of the autotuning approaches proposed in Refs.~\cite{Zwolak20-AQD, Durrer19-ATQ}.
    Panels (f)--(j) in the middle row show pixel classification results for the charge stability diagrams shown in the top row. 
    Plots of fitting results used to determine whether a spurious QD is present are shown in the bottom row [panels (k)--(o)].
    The different groups of transition are shown with different shades of green.
    The monotonicity within each group of transitions is clearly visible in panels (k) and (l).
    In contrast, in the three plots shown in panels (m)--(o), there is a clear divergence from the expected trend for the spurious QD, as indicated by the nonmonotonic change between the first (leftmost point) and second (middle point) transition.
    Error bars indicate one standard deviation.}
    \label{fig:unint_dot_example}
\end{figure*}

As a demonstration, we use the pixel classification and fit tools to analyze five experimental charge stability diagrams: two capturing properly formed QDs, shown in Figs.~\ref{fig:unint_dot_example}(a) and \ref{fig:unint_dot_example}(b), and three capturing spurious QDs, shown in Figs.~\ref{fig:unint_dot_example}(c)--\ref{fig:unint_dot_example}(e).
While for extraction of the virtualization matrix small scans are sufficient, detection of spurious QDs requires somewhat bigger scans to ensure that the neighboring charge transitions are adequately captured.
In our analysis, we rely on 2D scans of a size roughly three times the charging energy (four times the area of scans typically used in autotuning algorithms~\cite{Zwolak20-AQD, Durrer19-ATQ}).
We also consider only clusters consisting of at least $20$ pixels to ensure better reliability of the linear fit. 

After pixel classification, contiguous clusters of pixels belonging to a given class of transitions are analyzed individually, resulting in a cluster-based fit and standard deviation. 
Cases where more than one cluster belongs to a given charge transition result in separate fits, as in Figs.~\ref{fig:unint_dot_example}(b) and \ref{fig:unint_dot_example}(e) where the LT lines are split into groups to either side of the RT lines.
This separation serves two purposes: to ensure that variation along a given transition is not included and to treat each additional line independent of the charge on another QD.

Within a class and group of transitions, the magnitude of the capacitive coupling is expected to increase monotonically, while the spacing between consecutive transition lines decreases as a charge is added.
Such behavior is clearly visible in Figs.~\ref{fig:unint_dot_example}(k) and \ref{fig:unint_dot_example}(l), with the latter having to separate groups of fits (shown with different shades of green) for the groups of clusters. 
In contrast, a spurious QD can manifest itself by a nonmonotonic behavior of the capacitive coupling between transitions. 
This is depicted graphically by either the leftmost or the center point (or group of points) not following the expected decreasing trend in Figs.~\ref{fig:unint_dot_example}(m)--\ref{fig:unint_dot_example}(o).
The severity of this divergence can be quantified using a $Z$-test~\cite{Clog95-SMC}.

In practical applications, the automated detection of spurious QDs fits nicely within the autotuning paradigm. 
As mentioned earlier, many of the proposed approaches utilize a series of small 2D scans~\cite{Zwolak20-AQD, Durrer19-ATQ, Ziegler22-TRA, Ziegler22-TAR} or 1D rays~\cite{Zwolak20-RBC, Chatterjee21-AEC} as means to improve the tuning efficiency. 
While these approaches deliver measurement-cost-effective solutions, they are prone to unexpected and difficult-to-detect failure even when the data quality is high. 
Figures~\ref{fig:unint_dot_example}(b) and \ref{fig:unint_dot_example}(e) show examples of such potentially problematic cases.
The small 2D regions in the plunger-plunger space,  highlighted in these scans with the black boxes, are typical for topology setting algorithms.
In both cases, they are classified by a state classifier model as a double-QD state, with state prediction vectors being ${\rm\bf{p}}(V_{\mathcal{R}_{1
}})=[0.01, 0.04, 0.00, 0.04, 0.92]$ for region $V_{\mathcal{R}_{1
}}$ highlighted in Fig.~\ref{fig:unint_dot_example}(b) and ${\rm\bf{p}}(V_{\mathcal{R}_{2
}})=[0.00, 0.00, 0.18, 0.05, 0.76]$ for region $V_{\mathcal{R}_{2}}$ highlighted in Fig.~\ref{fig:unint_dot_example}(b), where ${\rm\bf{p}}(V_\mathcal{R})=[p_{\rm ND},\,p_{{\rm SD}_L},\,p_{{\rm SD}_C},\,p_{{\rm SD}_R},\,p_{\rm DD}]$ with ND denoting no QDs formed, ${\rm SD}_L$, ${\rm SD}_C$, and ${\rm SD}_R$ denoting the left, central, and right single QD respectively, and ${\rm DD}$ denoting the double-QD state.
Moreover, the data quality for these images is high in both cases, with ${\rm\bf{q}}(V_{\mathcal{R}_{1
}})=[1.0, 0.0, 0.0]$ for region  $V_{\mathcal{R}_{1}}$ and ${\rm\bf{q}}(V_{\mathcal{R}_{1}})=[0.99, 0.01, 0.00]$ for $V_{\mathcal{R}_{2}}$, where ${\rm\bf{q}}(V_\mathcal{R})=[p_{\rm high},\,p_{\rm mod},\,p_{\rm low}]$ with $p_{\rm high}$, $p_{\rm mod}$, and $p_{\rm low}$ denoting the probability of region $V_\mathcal{R}$ being assessed by the data quality control module as ``high,'' ``moderate,'' and ``low'' quality, respectively.
Thus, from the ML perspective, both these predictions are confidently correct.
However, when looked at within a slightly larger voltage range, it is clear that in the latter case the small scan captures an anticrossing with a spurious QD, which for practical tuning purposes is a failure. 
If not recognized and corrected for, termination at this point will result in an incorrect charge setting and virtualization~\cite{Durrer19-ATQ, Ziegler22-TAR}.

The spurious QD detection algorithm can be easily implemented in the autotuning algorithm proposed in Ref.~\cite{Ziegler22-TRA} as a safety check before the unloading step is initiated.
If a transition is flagged as potentially problematic, tuning may be suspended and measurements on other gates initiated to further investigate the problem.
Additional analysis to determine the separation between transitions can be implemented as a measure complementary to the monotonicity analysis. 
Automated identification and characterization of spurious QDs may also be useful to inform fabrication procedures and prevent them in future devices~\cite{Thorbeck12-UQD}.

\section{Conclusions}
\label{sec:conclusion}
As QD devices grow in size and complexity, the need for reliable and automated tune-up procedures becomes more pressing.
Establishing orthogonal control of the chemical potentials of QDs is one of the first steps in the tune-up of any larger QD array.
Here, we demonstrated a method that combines machine-learning-based pixel classification and traditional curve fitting to reliably determine voltage cross-talk coefficients. 
The advantage of this method over previous approaches is highlighted by increased reliability and resilience to experimental noise.
Further on, unwanted spurious dots that would reduce or inhibit device performance can be detected and flagged when this module is used as part of a larger tune-up algorithm~\cite{Ziegler22-TAR}.
The ability to automatically and reliably detect spurious QDs is especially important on wafer-scale fabrication characterization tools that produce more data than can efficiently be processed by human analysis.
In extensions, our tools could allow for automated navigation of voltage space for more targeted measurement of all chemical potential and tunnel barrier cross-capacitances~\cite{Hensgens17-FHQ, Hsiao20-EOT}.

\begin{acknowledgments}
This research was performed while J.Z. held an NRC Research Associateship award at the National Institute of Standards and Technology (NIST).
The views and conclusions contained in this paper are those of the authors and should not be interpreted as representing the official policies, either expressed or implied, of the U.S. Government. 
The U.S. Government is authorized to reproduce and distribute reprints for Government purposes notwithstanding any copyright noted herein. 
Any mention of commercial products is for information only; it does not imply recommendation or endorsement by NIST.
\end{acknowledgments}

\appendix
\section{The structure of the convolutional neural network}\label{app:cnn}
The convolutional neural network used in the pixel classifier model follows an architecture similar to a feature pyramid network~\cite{Lin17-FPN} with separable convolutions in the downsampling branch, residual connections, and batch normalization applied after each convolution and activation.
This structure is adapted from an example in Ref.~\cite{Chollet20-ISU}, with the major difference being long skip connections between the upsampling and downsampling branches of the network.
The downsampling branch has three pairs of convolutional layers of kernel size 3 and 16, 32, and 64 filters with residual connections between the input and output of each pair.
Residuals are computed using a $1\times1$ convolution followed by addition.
Each pair of convolutions is followed by a max pooling layer with a stride of 2.
The upsampling branch similarly has three pairs of transpose convolutional layers with 64, 32, and 16 filters, with residual connections between the input and output of each pair.
Each pair of convolutions is followed by an upsampling layer with a stride of 2.
Long skip connections go between the output of each downsampling layer to the output of each upsampling layer with a matching number of the image size using a convolutional layer of kernel size 3.
The CNNs are trained using the Adam optimizer~\cite{Kingma14-ASO} with a learning rate of $1\times10^{-3}$.

%

\end{document}